# Graphene Induced Large Shift of Surface Plasmon Resonances of Gold Films: Effective Medium Theory for Atomically Thin Materials

Md Kamrul Alam[1], Chao Niu[2], Yanan Wang[3,4], Wei Wang[5,6], Yang Li[4], Chong Dai[7], Tian Tong[4] Xiaonan Shan[4], Earl Charlson[4], Steven Pei[4], Xiang-Tian Kong[3], Yandi Hu[7], Alexey Belyanin[8], Gila Stein[9], Zhaoping Liu[5,6], Jonathan Hu[2*], Zhiming Wang[3,10] and Jiming Bao[1,4*]

[1]*Materials Science and Engineering*
*University of Houston*
*Houston, Texas 77204, USA*

[2]*Department of Electrical & Computer Engineering*
*Baylor University*
*Waco, TX 76798, USA*

[3]*Institute of Fundamental and Frontier Sciences*
*University of Electronic Science and Technology of China*
*Chengdu, Sichuan 610054, China*

[4]*Department of Electrical and Computer Engineering*
*University of Houston*
*Houston, TX 77204, USA*

[5]*Ningbo Institute of Materials Technology & Engineering*
*Chinese Academy of Sciences*
*Ningbo, Zhejiang 315201, China*

[6]*Key Laboratory of Graphene Technologies and Applications of Zhejiang Province*
*Ningbo Institute of Materials Technology and Engineering, Chinese Academy of Sciences*
*Ningbo, Zhejiang 315201, China*

[7]*Department of Civil and Environmental Engineering*
*University of Houston*
*Houston, TX 77204, USA*

[8]*Department of Physics & Astronomy*
*Texas A&M University, College Station, TX 77843, USA*

[9]*Chemical and Biomolecular Engineering*
*The University of Tennessee*
*Knoxville, TN 37996, USA*

[10]*State Key Laboratory of Electronic Thin Films and Integrated Devices*
*University of Electronic Science and Technology of China*
*Chengdu, Sichuan 610054, China*

Abstract

Despite successful modeling of graphene as a 0.34-nm thick optical film synthesized by exfoliation or chemical vapor deposition (CVD), graphene induced shift of surface plasmon resonance (SPR) of gold films has remained controversial. Here we report the resolution of this controversy by developing a clean CVD graphene transfer method and extending Maxwell-Garnet effective medium theory (EMT) to 2D materials. A SPR shift of 0.24º is obtained and it agrees well with 2D EMT in which wrinkled graphene is treated as a 3-nm graphene/air layered composite, in agreement with the average roughness measured by atomic force microscope. Because the anisotropic built-in boundary condition of 2D EMT is compatible with graphene's optical anisotropy, graphene can be modelled as a film thicker than 0.34-nm without changing its optical property; however, its actual roughness, *i.e.*, effective thickness will significantly alter its response to strong out-of-plane fields, leading to a larger SPR shift.

# INTRODUCTION

Maxwell-Garnet effective medium theory (EMT) was developed more than 100 years ago to obtain the macroscopic dielectric property of an inhomogeneous medium[1, 2]. The Maxwell-Garnet (M-G) mixing formula provides us the permittivity of a composite in terms of the permittivity and volume fraction of the individual constituents in a host medium[1-3]. The theory becomes more important today as nanostructures and nanomaterials are routinely synthesized and assembled to make nanocomposites or metamaterials for the desired electromagnetic responses and functionalities. Because the original mixing formula is based on non-interacting spherical inclusions in a host medium, it has been revised to handle non-spherical inclusions with mutual interaction[3-15]. The original and revised mixing formulas have been proven to be powerful tools in accurately capturing the macroscopic electromagnetic responses of composite materials, and good agreements have been demonstrated between theory and experiment for many systems such as metal-ceramic films[6, 16], polymer-ceramic composites[17], amorphous silicon thin films[18], polymer-single-walled carbon nanotube composite[8], and aligned carbon nanotube film [19, 20]. However, all these studies only investigated one or three dimensional structures in three-dimensional host media, EMT for two dimensional (2D) layered structures have not been evaluated although the theory was developed long ago[21] and atomically thin 2D structures have become widely available.

Graphene, a truly atomically thin nanomaterial, has been treated as a 3D-like flat thin film with $n$ and $k$, real and imaginary parts of the refractive index, and with a finite thickness since its first optical characterization using spectroscopic ellipsometry[22]. Its picture as a 0.34-nm film, no matter if it is exfoliated or grown by chemical vapor deposition (CVD), has worked very well in nearly all optical characterizations such as ellipsometry[22-27], attenuated total reflection (ATR)[28, 29] and reflection spectroscopy[30, 31]. Surface plasmon resonance (SPR) of gold film in the Kretschmann configuration is sensitive to minute changes on a sample's surface, so it is an ideal tool to explore basic optoelectronic property of thin dielectric films and study their light-matter interactions. However, there has been a big discrepancy between theory and experiment. Based on the flat graphene picture, the SPR shift of an Au film in resonant angle with and without single layer graphene in air is calculated to be less than 0.1 degree[32-38].

Experimentally, except for micrometer-size exfoliated graphene[39], SPR shift induced by large-size CVD graphene is more than twice the calculated value, varying from 0.24 to 1 degree [15, 40-42].

In this work, we apply an effective media theory to atomically thin material and report the resolution of graphene SPR puzzle both experimentally and theoretically. We first develop a polymer-free CVD graphene transfer method to make sure that SPR shift is induced by graphene only. We then point out several mistreatments in previous calculations, and a good agreement is achieved using actual roughness of graphene and 2D EMT: atomically thin materials should be treated as a flat film with effective thickness depending on its intrinsic surface roughness. Finally, we show that the Kretschmann configuration is an excellent platform to test 2D EMT and characterize anisotropic 2D composite films.

**RESULTS AND DISCUSSION**

Monolayer graphene was grown on polycrystalline Cu foils using a home build CVD system[43]. In order to avoid any potential chemical contamination[44, 45], we developed a PMMA-free graphene transfer technique. Our approach took advantage of the hydrophobic nature of graphene, which makes graphene float on the etching solution without any polymer supporting layer[45]. Figure 1a shows a floating graphene/copper on aqueous iron nitrate etching solution. Red marks were placed on the corner of the graphene before etching to make it visible. After Cu was etched out, DI water was slowly added from the top, and the etching solution was drained from the bottom until it was completely removed (Figures 1b-c). 100 mL of hydrochloric acid (HCl 5 M) was finally injected into the container and then replaced by DI water again to eliminate Fe residues. A floating graphene on DI water was scooped by Au/glass substrate and could be transferred to any substrates in principle. This method is simpler than many reported polymer-free or support-free methods[46-51]. High-quality monolayer graphene was confirmed by Raman (Fig. S1) and optical transmission (Fig. S2)[52, 53].

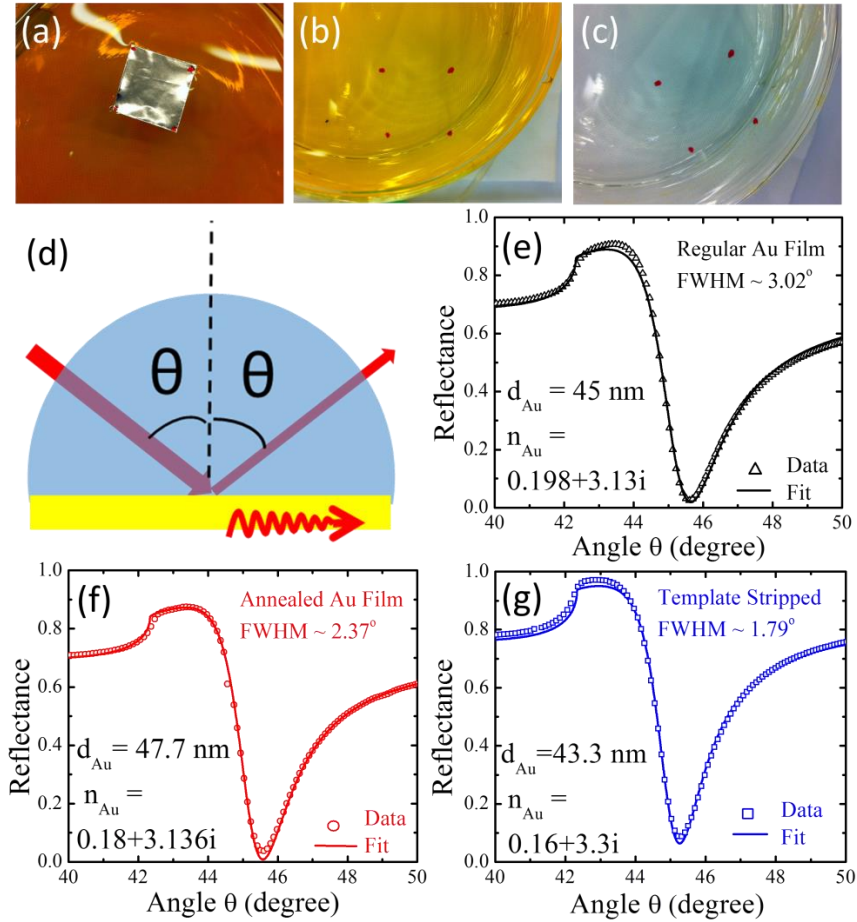

Figure 1. PMMA-free Graphene transfer and SPR of three Au films. (a) Graphene covered Cu foil floating on iron nitrate etching solution. (b-c) Graphene floats on (b) etching solution and eventually on (c) DI water after gradual solution replacement. (d) Schematic of Kretschmann configuration with a hemisphere prism and Au film. (e-g) Surface plasmon resonance (SPR) curves of three Au films and corresponding fitting curves and parameters. $n_{glass}$ = 1.485. $n_{Au}$ and $d_{Au}$ are index and thickness of the Au film, respectively.

Three types of Au films were first prepared and characterized with SPR before graphene transfer: regular Au film fabricated with electron-beam evaporation, regular film after thermal annealing, and template stripped gold (TSG) film[54]. They all had a nominal thickness of 45 nm and were evaporated (regular Au and annealed) or attached (TSG) to glass slides which are index matched with that of the SPR prism. Fig. 1d shows the schematic of the Kretschmann configuration. A 633-nm HeNe laser was used to excite surface plasmon. Figs. 1e-g show SPR curves of three Au films in ambient air. All of them display characteristic features with the minimum reflectance around 45.5°, indicating the excitation of SPR. These SPR curves can be well fitted by treating Au film as a homogenous layer with adjustable index[55-58]. The obtained refractive index (RI) of the Au films are included in the figures, they are close to each other and fall within the typical index range of Au[58-60], indicating a successful modeling of Au film SPR.

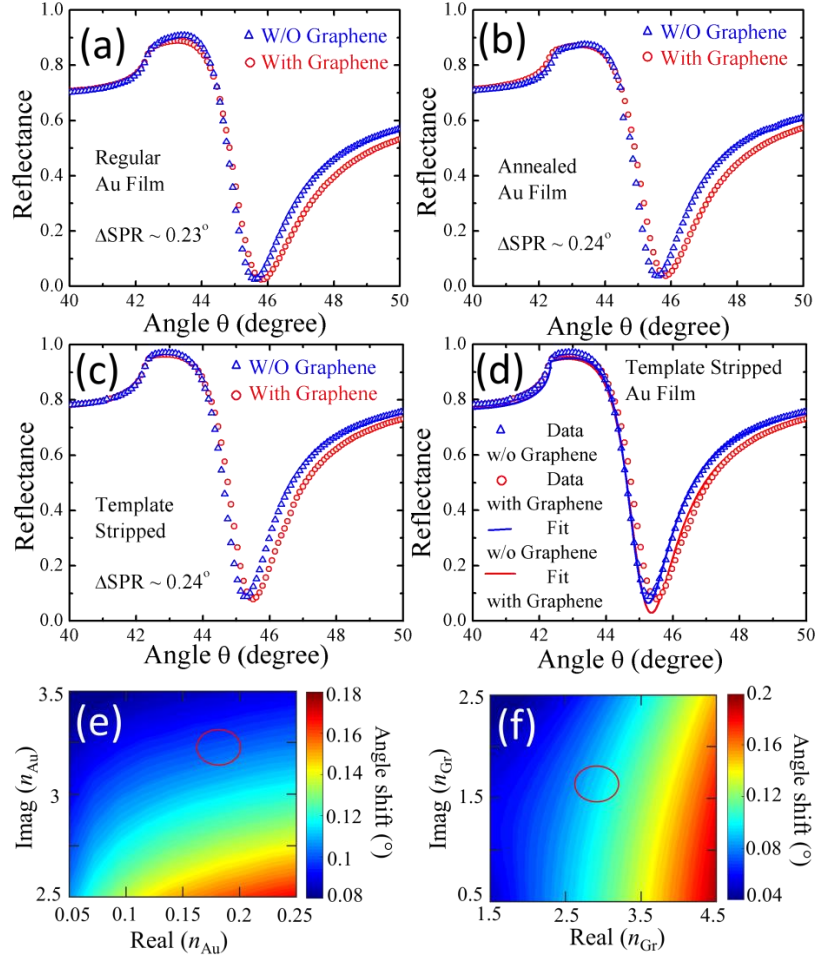

Figure 2. Graphene-induced SPR shifts and simulations with variable Au and graphene index. (a-c) SPR shifts of the three Au films induced by monolayer graphene. (d) Measured and fitted SPR curves of TSG films (red) with and (blue) without graphene. Effect of refractive index of (e) Au and (f) graphene on the SPR shift. For (e), the index and thickness of graphene is fixed as $n_{Graphene}=2.95+1.54i$. For (f), the index and thickness of Au film is chosen to be $0.18+3.30i$ and 45 nm.

A major advantage of SPR is that it is very sensitive to the dielectric environment of the Au film. Figs. 2a-c show SPR curves for the same three films after transferring monolayer graphene. For comparison, the initial curves without graphene are also included. It can be seen that graphene does have induced a significant change to each SPR curve with a similar SPR shift of ~0.24° despite different surface roughness of three Au films. Similar measurements have been reported, but our shift is among the lowest with CVD graphene[15, 40-42]. We believe this is due to our PMMA-free graphene transfer technique since any additional contamination will increase SPR shift.

Graphene induced SPR shift has also been calculated by many groups[32-38]. However, the calculated values are less than half of the lowest experimental value[15, 40-42]. The predicted shift of ~0.1 degree is always reached when graphene is treated as a 0.34-nm thick flat homogeneous film with a refractive index of ~2.95+1.54i, which is an in-plane index obtained by ellipsometry[22, 32-38]. The same SPR shift is obtained for our three types of Au films if we follow the same modelling approach. In fact, such a small SPR shift will always be obtained within experimental variation of index with Au films and graphene. To prove this point, we calculate the SPR shift as functions of real and imaginary parts of the refractive indices of both graphene and Au film. Figs. 2e-f show that the SPR shift varies smoothly and there is no hot spot with an abrupt large shift. For Au film, a change of 10% in the imaginary part has a larger effect than the change in real part, but its imaginary part has very little variation among different Au films. For typical index range of Au film and graphene (red circles on the images), SPR shift is found to be more or less around a small value of 0.10°.

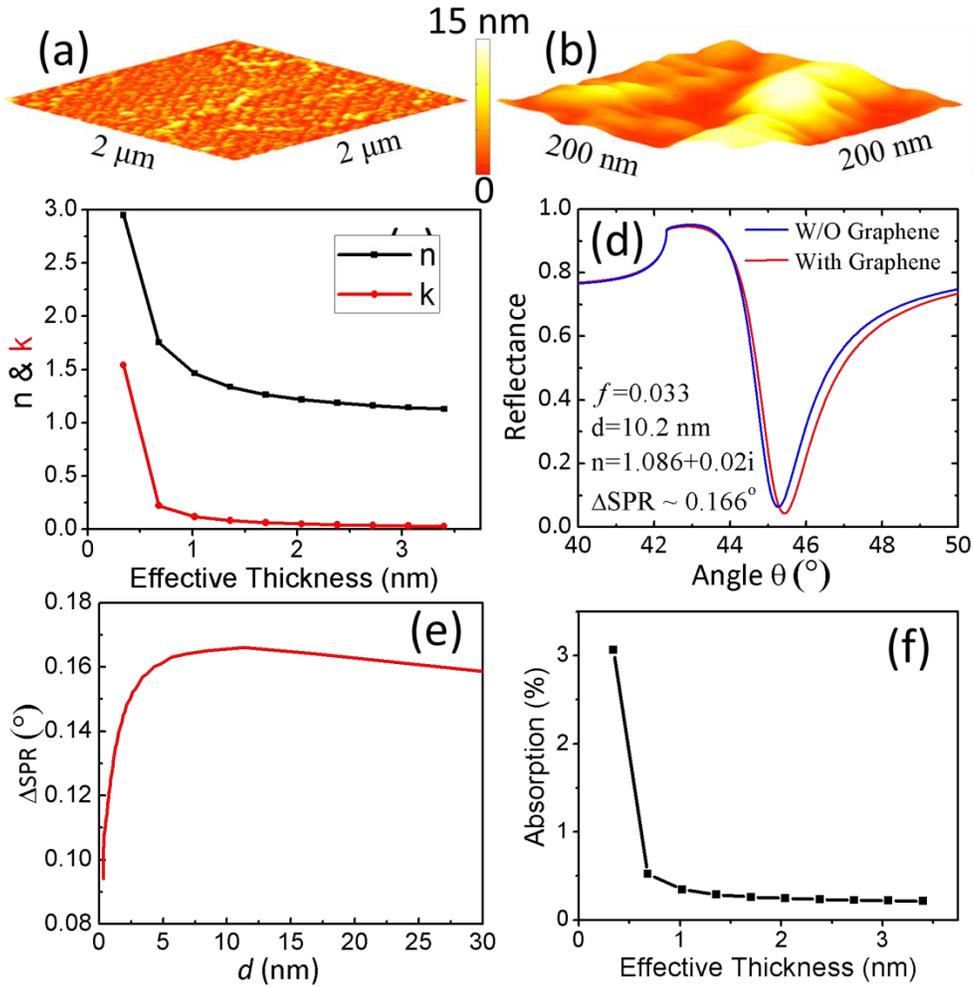

Figure 3. Graphene AFM images and Maxwell-Garnett EMT. (a) 3D AFM image of graphene with a 2×2 μm² scan area on the TSG Au substrate. Root mean square (RMS) roughness is 3.02 nm. (b) Zoomed-in image of the roughest center region in (a). (c) n and k of graphene/air composite as a function of effective thickness d. (d) Representative example of SPR curves with and without graphene/air composite film with an effective thickness d of 10.2 nm. (e) Calculated SPR shift as a function of effective thickness $d$ of graphene/air composite. (f) Thickness-dependent optical absorbance of graphene/air composite calculated by $A = \frac{4\pi nkd}{\lambda}$.

This persistent disagreement on SPR shift between theory and experiment has existed for quite a time, but no serious attention has been paid to this issue. Obviously, the model of flat homogeneous graphene is oversimplified because graphene, especially grown by CVD is not perfectly flat microscopically but tends to form wrinkles, nano ripples and corrugations with surface roughness ranging from few to ten nanometers[61-70]. This can be seen from a representative AFM image in Figs. 3a-b. Nevertheless, graphene is still well aligned in the same plane with out-of-plane tilt angle less than ~5° for more than 90% of total area. If we approximate graphene as a rippled sheet embedded in a thin layer air, then we can use the M-G mixing formula to calculate its effective $n$ and $k$ and then use them to calculate the SPR shift. The actual effective thickness $d$ of graphene/air can be estimated from AFM image, and it determines the graphene volume fraction or filling factor. Because graphene is relatively flat, it is safe to assume that the total amount of graphene is the same as monolayer. The volume fraction $f = t/d$, where $t = 0.34$ nm is the thickness of monolayer graphene and $d$ is the effective thickness. The dielectric constant of the composite will be given by[3]

$$\varepsilon_{eff} = \frac{1+2f\frac{\varepsilon-1}{\varepsilon+2}}{1-f\frac{\varepsilon-1}{\varepsilon+2}} \qquad (1)$$

where $\varepsilon = (2.95 + 1.54i)^2 = 6.33 + 9.09i$. Fig. 3c plots the thickness dependent effective $n$ and $k$ of graphene/air based on Eq. (1). Both $n$ and $k$ decreases quickly as $d$ increases, while k decreases much faster than $n$, dropping by more than a half when the $d$ doubles. Fig. 3d shows an example of SPR curve when $d$ is 10.2 nm. A SPR shift of 0.17° is obtained. This shift is much larger than that with the flat graphene, however, when we calculate the thickness dependent SPR shift as shown in Fig. 3e, we found a maximum shift of 0.17° regardless the roughness of graphene. As $d$ goes beyond 10 nm, the SPR shift starts to decrease, making it impossible to reach 0.24°. This failure of M-G theory can also be seen from the sharp decrease in optical absorption of the graphene/air composite in Fig. 3f. Experimentally, all the graphene, no matter it is grown by exfoliation or CVD, the transmission is kept at ~97%[71, 72]. This failure of M-G EMT is anticipated because graphene is not a spherical inclusion, and graphene's dielectric constant is intrinsically not isotropic[22].

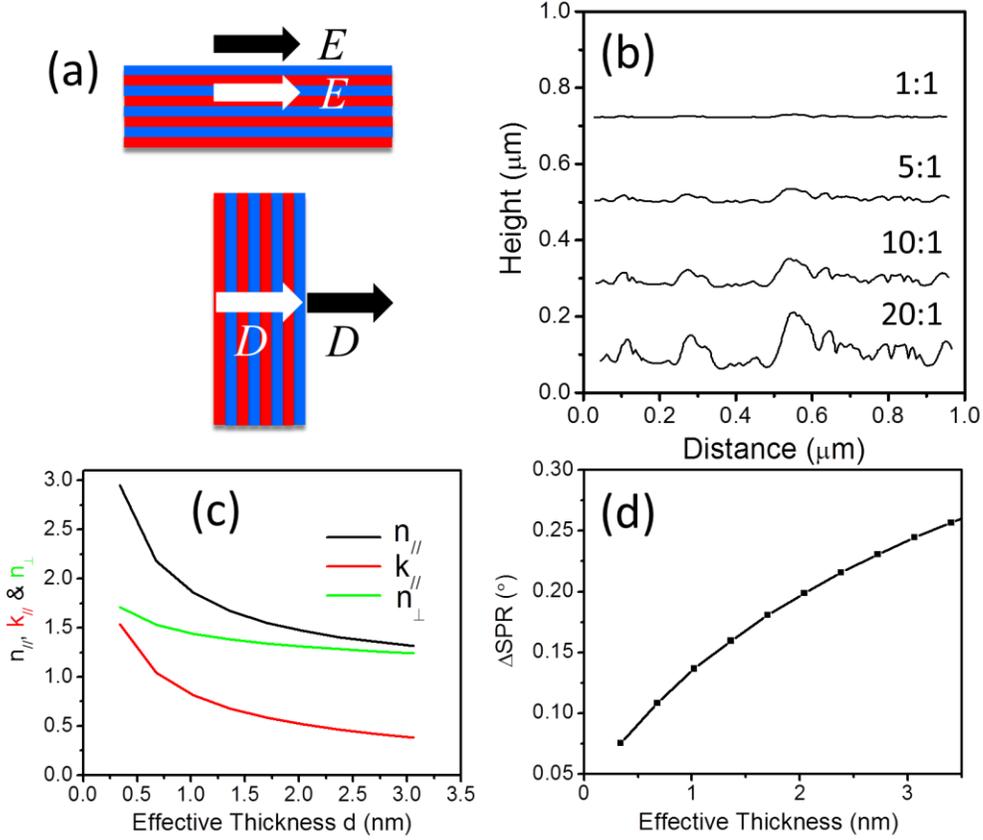

Figure 4. Effective medium theory for 2D layered composite and its application to graphene SPR shift. (a) Boundary condition. The tangential component of the electric field *E* and the normal component of the displacement *D* are continuous. (b) Line scan across the center roughest region of the AFM in Fig. 3a-b at different height to distance scale. (c) *n* and *k* of graphene/air composite as a function of effective thickness *d*. (d) Calculated SPR shifts as a function of effective thickness *d*.

Since graphene is nearly parallel to the Au film, we can approximate it as a graphene/air layered composite. Such layered or stratified composite was initially studied theoretically by Rytov in 1956[21]. Unlike isotropic spherical inclusions, Figs. 4a show anisotropic boundary condition for *E* and *D*, indicating that 2D layered composite is intrinsically anisotropic even if each constituent material is isotropic. It is important to point out that graphene is still quite flat (Fig. 4b), it is also anisotropic, although it has been treated as an isotropic medium in previous calculations of SPR shift[32-38]. The effective dielectric constant can be conveniently derived based on its definition and Maxwell equation's boundary conditions: $\varepsilon_{eff}$ is the ratio of average electric flux density, $D$, to average electric field, $E$; $E$ is continuous in parallel direction) and $D$ is continuous in perpendicular direction[3].

In parallel direction, for two layered materials with volume fractions of $f_a$ and $f_b$, the average $D_\parallel$ is given by

$$D_\parallel = f_a D_{\parallel a} + f_b D_{\parallel b} = f_a \varepsilon_{\parallel a} E_{\parallel a} + f_b \varepsilon_{\parallel b} E_{\parallel b} = (f_a \varepsilon_{\parallel a} + f_b \varepsilon_{\parallel b}) E_\parallel = \epsilon_\parallel E_\parallel$$

Since $E_{\parallel a} = E_{\parallel b} = E_\parallel$, we obtain

$$\epsilon_\parallel = f_a \varepsilon_{\parallel a} + f_b \varepsilon_{\parallel b} \tag{2}$$

In perpendicular direction, we have average electrical field $E_\perp$, which is given by

$$E_\perp = f_a E_{\perp a} + f_b E_{\perp b} = f_a \frac{D_{\perp a}}{\varepsilon_{\perp a}} + f_b \frac{D_{\perp b}}{\varepsilon_{\perp b}} = \left(\frac{f_a}{\varepsilon_{\perp a}} + \frac{f_b}{\varepsilon_{\perp b}}\right) D_\perp = \frac{D_\perp}{\varepsilon_\perp}$$

Since $D_{\perp a} = D_{\perp b} = D_\perp$, we obtain

$$\frac{1}{\varepsilon_\perp} = \frac{f_a}{\varepsilon_{\perp a}} + \frac{f_b}{\varepsilon_{\perp b}} \tag{3}$$

The volume fraction of graphene is still $0.34/d$. Fig. 4c shows the effective index of $n$ and $k$ in both parallel and perpendicular directions as d increases. Note that because the out-of-plane optical absorption of single layer graphene is zero[22], effective $k$ in the perpendicular direction is also zero. Fig. 4d shows the effect of effective thickness $d$ on the SPR shift. Because the index decreases much slower than in the 3D EMT case, a larger shift is achieved. Based on the SPR shift, the effective thickness should be around 3 nm, which agrees with the average roughness of 3.02 nm calculated from the AFM image.

The success of 2D EMT can also be verified by many far-field optical observations of graphene. Because imaginary part of air's dielectric constant is zero, Eq. (2) can be written as

$$\text{Im}(\epsilon_\parallel) = f\text{Im}(\varepsilon_{\parallel a}) = \frac{t}{d}\text{Im}(\varepsilon_{\parallel a}), \tag{4}$$

where Im stands for the imaginary part of the variable. Since $\varepsilon = \varepsilon_R + i\varepsilon_I = (n + ik)^2$, we have $\varepsilon_I = 2nk$. Thus Eq. (4) can be written as

$$2n_\| k_\| = \frac{t}{d} 2n_{\|g} k_{\|g}, \text{ i.e. } dn_\| k_\| = t n_{\|g} k_{\|g} \tag{5}$$

Eq. (5) shows that $dn_\| k_\|$ is a constant regardless of the effective thickness $d$. According to Beer's law, the intensity of light travel through a thin film goes as $I(x) = I_o e^{-\frac{4\pi n k x}{\lambda}}$, so $dn_\| k_\|$ determines the optical absorption of light through the film with a thickness $d$. In other words, absorption for the normal incident light through graphene is a constant no matter graphene is flat or corrugated. This conclusion has been verified by our optical transmission spectrum (Fig. S2) and numerous other experimental observations[71, 73-76].

It is not surprising that a thicker graphene/air composite film from 2D EMT can have the same optical absorption as that of an original thin flat graphene. We further argue that this treatment of rough graphene does not affect any of its optical properties in conventional thin film optical characterizations when it is surrounded by dielectric media. This can be understood as follows. In principle graphene is an atomic network of carbon atoms, it should be treated as an infinitely thin sheet. A finite thickness of 0.34 nm is only a convenient choice, it can be varied in ellipsometry as long as *n* and *k* are also adjusted accordingly to fit the data. In this sense, original picture of a flat 0.34 nm thick graphene is already an approximation. When we further increase its thickness using 2D effective medium theory, we have kept **E** and **D** boundary condition the same as before. For conventional thin film optical characterizations, as long as the effective thickness is much smaller than the wavelength of light, the results remain the same.

However, the above argument becomes invalid when a graphene is placed on the surface of a plasmonic or metallic film due to the following two reasons. First, the electrical field near the surface called near field does not remain constant as when graphene is surrounded by a dielectric media; instead, it changes rapidly over a short distance above the surface. Second, the near field is dominated by the field normal to the surface or graphene. These unique differences can be seen in Fig. 5a when surface plasmon is excited by the incident laser. The color indicates the normalized intensity of the electric field and the arrows indicates the direction of the electric field. As a result, the out-of-the-plane dielectric constant of graphene matters, and previous calculations based on in-plane dielectric constant are not accurate; the effective thickness is also important: thicker effective layer increases graphene interaction with normal near field, leading to a larger SPR shift. To demonstrate this point, we calculate SPR shifts as a function of the effective thickness in isotropic and anisotropic cases. Fig. 5b shows that the difference increases as the thickness increases. For isotropic treatment, the in-plane optical constant from 2D EMT is used as both in-plane and out-plane constants. The isotropic calculation clearly overestimates the shift because the out-plane index is much smaller than the in-plane optical constant. Certainly, this treatment of isotropic 2D media is not self-consistent. Note that for a large effective thickness, 2D EMT also becomes invalid because the composite cannot be approximated as a layered structure defined Fig. 4a.

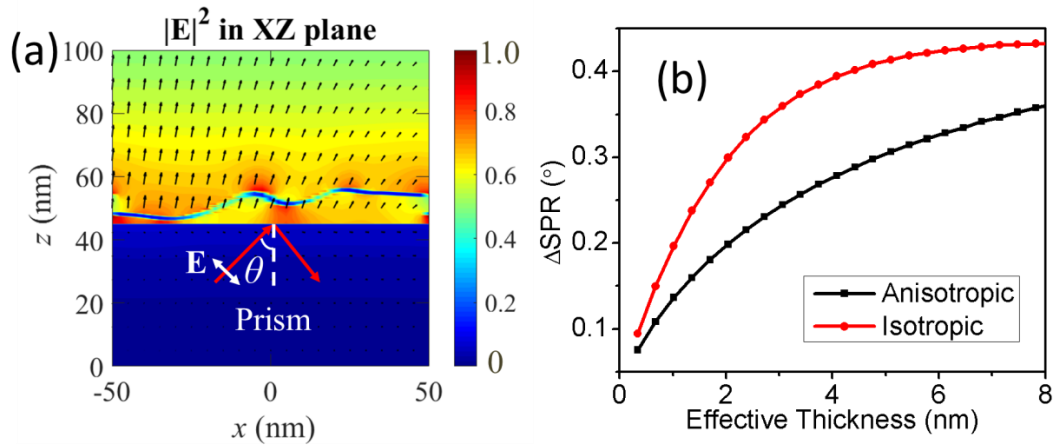

Figure 5. The effect of perpendicular field and out of plane index on the SPR shift. (a) Cross-sectional view of electric field near the Au surface at a critical incident angle of 45.5°. The wrinkled line is graphene from AFM images in Fig. 3a-b. (b) Thickness dependent SPR shifts when graphene/air composite is treated as an isotropic or anisotropic media.

**CONCLUSIONS**

In summary, we have successfully extended traditional M-G mixing theory for 3D isotropic media to 2D layered structures and applied 2D EMT to graphene. A good agreement of graphene-induced SPR shift between theory and experiment is achieved after wrinkled graphene is treated as an anisotropic graphene/air layered composite. We also revealed a unique property of 2D EMT: normal incident optical absorption and typical optical properties remain the same regardless the effective thickness of layered composite. We point out that previous treatments of graphene as an isotropic medium is not accurate, and that the Kretschmann configuration is an excellent platform to measure anisotropic optical constant of 2D material and test 2D effective medium theory due to its strong normal near field on the surface. This picture of graphene as an effective medium is applicable to other atomically thin nanomaterials or layered structure such as such as graphene oxide (GO), reduced GO, transition metal dichalcogenides, 2D material-based nanocomposite or metamaterials, and helps to understand their electromagnetic responses and functionalities such as enhanced SPR sensitivity[33, 34, 77-83].